\documentclass[preprint,aps,showpacs,showkeys,preprintnumbers,superscriptaddress,amsmath,amssymb]{revtex4-1}

\usepackage{graphicx}
\usepackage{siunitx}
\usepackage{color}

\newcommand{\NbFey}{Nb$_{1-y}$Fe$_{2+y}$}
\newcommand{\NbFeFe}{Nb$_{0.985}$Fe$_{2.015}$}
\newcommand{\NbFeNb}{Nb$_{1.01}$Fe$_{1.99}$}
\newcommand{\NbFe}{NbFe$_2$}
\newcommand{\YRS}{YbRh$_2$Si$_2$}
\newcommand{\Hstar}{\ensuremath{H^{\star}}}
\newcommand{\Tstar}{\ensuremath{T^{\star}}}
\newcommand{\astar}{\ensuremath{a^{\star}}}
\newcommand{\bstar}{\ensuremath{b^{\star}}}
\newcommand{\TFM}{\ensuremath{T_{\text{c}}}} 		
\newcommand{\TAFM}{\ensuremath{T_{\text{N}}}}	
\newcommand{\TC}{\ensuremath{T_{\text{0}}}}			
\newcommand{\HAFM}{\ensuremath{H_{\text{N}}}}	

\newcommand{\dd}{\text{d}}																
\preprint{ver 10.0 confidential}
\newcommand{\sven}[1]{
{#1}}

\begin{document}



\title{Quantum Tricritical Points in \NbFe}

\author{Sven Friedemann}
\affiliation{HH Wills Laboratory, University of Bristol, Bristol BS8 1TL, UK}
\affiliation{Cavendish Laboratory, University of Cambridge, Cambridge CB3 0HE, UK}

\author {Will J.~Duncan}
\affiliation{Department of Physics, Royal Holloway, University of London, Egham TW20 0EX, UK}

\author{Max Hirschberger}
\altaffiliation{Current address: Department of Physics, Princeton University, Jadwin Hall, Princeton NJ 08544, USA}
\affiliation{Cavendish Laboratory, University of Cambridge, Cambridge CB3 0HE, UK}
\affiliation{Physik Department, Technische Universit\"at M\"unchen,
  James Franck Stra\ss e, 85748 Garching, Germany}

\author{Thomas Bauer}
\affiliation{MPI-CPfS, N\"othnitzer Strasse, 01189 Dresden, Germany}

\author{Robert K\"uchler}
\affiliation{MPI-CPfS, N\"othnitzer Strasse, 01189 Dresden, Germany}

\author{Andreas Neubauer} 
\affiliation{Physik Department, Technische Universit\"at M\"unchen,
  James Franck Stra\ss e, 85748 Garching, Germany}

\author{Manuel Brando}
\affiliation{MPI-CPfS, N\"othnitzer Strasse, 01189 Dresden, Germany}

\author {Christian Pfleiderer}
\affiliation{Physik Department E21, Technische Universit\"at M\"unchen,
  James Franck Stra\ss e, 85748 Garching, Germany}

\author{F.~Malte Grosche}
\affiliation{Cavendish Laboratory, University of Cambridge, Cambridge CB3 0HE, UK}

\date{\today}


\pacs{
75.30.Fv,
75.30.Kz,
74.40.Kb
}
\begin{abstract}
Quantum critical points (QCPs) emerge when a 2nd order phase transition is suppressed to zero temperature. In metals the quantum fluctuations at such a QCP can give rise to new phases including unconventional superconductivity. Whereas antiferromagnetic QCPs have been studied in considerable detail ferromagnetic (FM) QCPs are much harder to access\cite{Loehneysen2007,Brando2016}. In almost all metals FM QCPs are avoided through either a change to 1st order transitions or through an intervening spin-density-wave (SDW) phase. Here, we study the prototype of the second case, \NbFe. We demonstrate that the phase diagram can be modelled using a two-order-parameter theory in which the putative FM QCP is buried within a SDW phase. 
We establish the presence of quantum tricritical points (QTCPs) at which both the uniform and finite $q$ susceptibility diverge. The universal nature of our model suggests that such QTCPs arise naturally from the interplay between SDW and FM order and exist generally near a buried FM QCP of this type. Our results promote \NbFe\ as the first example of a QTCP, which has been proposed as a key concept in a range of narrow-band metals, including the prominent heavy-fermion compound \YRS\ 
\cite{Misawa2008}.
\end{abstract}

\maketitle

\noindent

Transition metal compounds with low-temperature magnetic order offer
attractive departure points in the study of correlated electron
materials. Key materials such as MnSi, ZrZn$_2$ or Ni$_3$Al have been
investigated for many years; high quality and well characterized
single crystals are widely available, their magnetic states have been
studied in detail, the magnetic excitation spectrum and their
electronic structure are often known from neutron scattering and
quantum oscillation measurements. A semi-quantitative
understanding of key properties such as the size of the ordered
moment, the ordering temperature and the low temperature heat capacity
is achieved within spin fluctuation theory
\cite{moriya85,lonzarich97}. Close to the border of
magnetism, however, the predictions of conventional spin fluctuation
theory no longer apply providing a long-standing
fundamental challenge to our understanding of correlated electron
systems \cite{Brando2016}. Key discrepancies concern firstly, the low temperature
form of the electrical resistivity $\rho(T)$, which follows a still
insufficiently understood $T^{3/2}$
power-law temperature dependence on the paramagnetic side of the FM quantum phase transition
\cite{pfleiderer01,takashima07,smith08, brando08} and secondly, the fate
of ferromagnetic (FM) order itself: rather than being continuously
suppressed towards a FM quantum critical point (QCP), the FM QCP is
avoided in clean metals.

One scenario for the avoidance, by which the FM transition becomes 1st
order near the putative QCP, is well understood by theory and
experimentally well established
\cite{belitz99,pfleiderer94,Pfleiderer97,Uhlarz04}.  The alternative
scenario, namely that the FM QCP is masked -- or buried -- by emergent
modulated magnetic order, has been discussed theoretically
\cite{belitz97,chubukov04,Conduit2009,Pedder2013}, but with the
exception of early work on \NbFe\ \cite{brando08,moroni09} and
recent studies on LaCrGe$_3$ \cite{Taufour16} as well as local moment systems PrPtAl
\cite{Abdul-Jabbar2015}, CeRuPO \cite{Kotegawa13}, CeFePO \cite{Lausberg12}, and YbRh$_2$Si$_2$ \cite{Lausberg13} this second scenario has so far been little explored experimentally.
\sven{Many of these materials bear the complication 
of additional energy scales from interactions 
between conduction electrons, localised 
$f$-electrons, crystal field levels, and complicated magnetic order. 
The transition metal itinerant magnet \NbFe\ avoids these complications
and has a simple crystal and magnetic structure.}


NbFe$_2$ can be tuned across a FM quantum phase transition by slight changes
in the composition that preserve good crystal quality, enabling
multi-probe studies without the complications of high pressure \cite{brando08}.
Whereas Nb$_{1-y}$Fe$_{2+y}$ orders ferromagnetically at
low temperature for modest levels of iron excess $y>\SI{0.01}{}$, 
stoichiometric or Nb-rich NbFe$_2$ has long been known to exhibit
signatures of a further magnetic phase transition
\cite{shiga87,yamada88,crook95,moroni09}, which has recently been
proven by microscopic probes to tip the system into long-wavelength
SDW order \cite{rauch15,niklowitz2017}.  The SDW transition temperature
itself extrapolates smoothly to zero for $y \simeq -\SI{0.015}{}$, and
near this SDW QCP, the heat capacity Sommerfeld coefficient exhibits a
logarithmic temperature dependence, whereas $\rho(T)$ follows a
$T^{3/2}$ power-law form \cite{brando08}.  These prior findings
strongly support the long-standing proposal \cite{moroni09} that a FM
QCP is indeed buried within an emergent SDW phase in NbFe$_2$ and
motivate a closer investigation. Here, we present detailed
magnetic, electric transport and thermal expansion data collected in
newly available high quality single crystals of the
Nb$_{1-y}$Fe$_{2+y}$ system at key compositions in the phase diagram. We
show that our data are consistent with a two-order-parameter
Landau theory \cite{moriya77}, which provides a novel and convenient
framework for extracting robust findings, namely that (i) the avoided
FM QCP can be located accurately inside the emergent SDW dome, (ii)
the presence of SDW order causes the FM transition to become first
order, and (iii) quantum tricritical points (QTCPs)
emerge at finite field. Both SDW and FM fluctuations associated
with the FM QTCPs will contribute to the excitation spectrum near the SDW
QCP, which may explain the seemingly contradictory temperature
dependencies of the heat capacity and resistivity in NbFe$_2$ mentioned above.
Thus our results provide new routes towards understanding the 
enigmatic physics of materials at the border of ferromagnetism.

\section{Results}

\begin{figure}[t]
\includegraphics[width=\columnwidth]{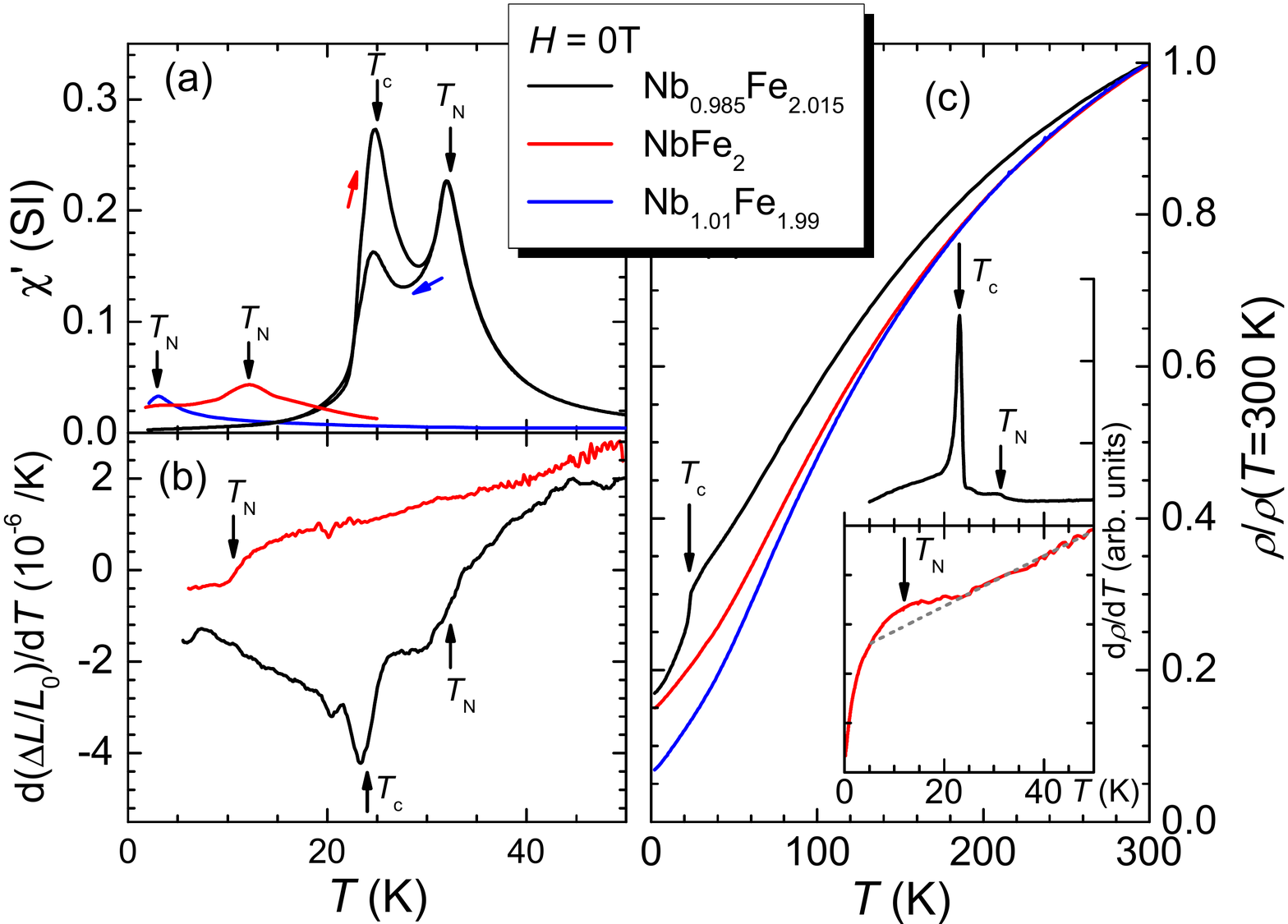}  
  \caption{
		\label{ChiRes} 
		Temperature dependence of the c-axis AC magnetic
    susceptibility $\chi(T)$ (a), the c-axis 
		linear thermal expansion $\dd L/(L_0 \dd T)$ (b), 
		and of the electrical resistivity $\rho(T)$ (c) 
		and its temperature derivative $\dd\rho/\dd T$ 
		(inset in c)  for \NbFey\ with $y=\SI{0.015}{}$ and $y=0$.
		Vertical black arrows indicate the transitions 
		at \TFM\ and \TAFM. 
		Red and blue arrows in (a) indicate the measurement 
		taken on warming and cooling.
	}
\end{figure}

\begin{figure*}%
\centering
	\includegraphics[width=.8\textwidth]{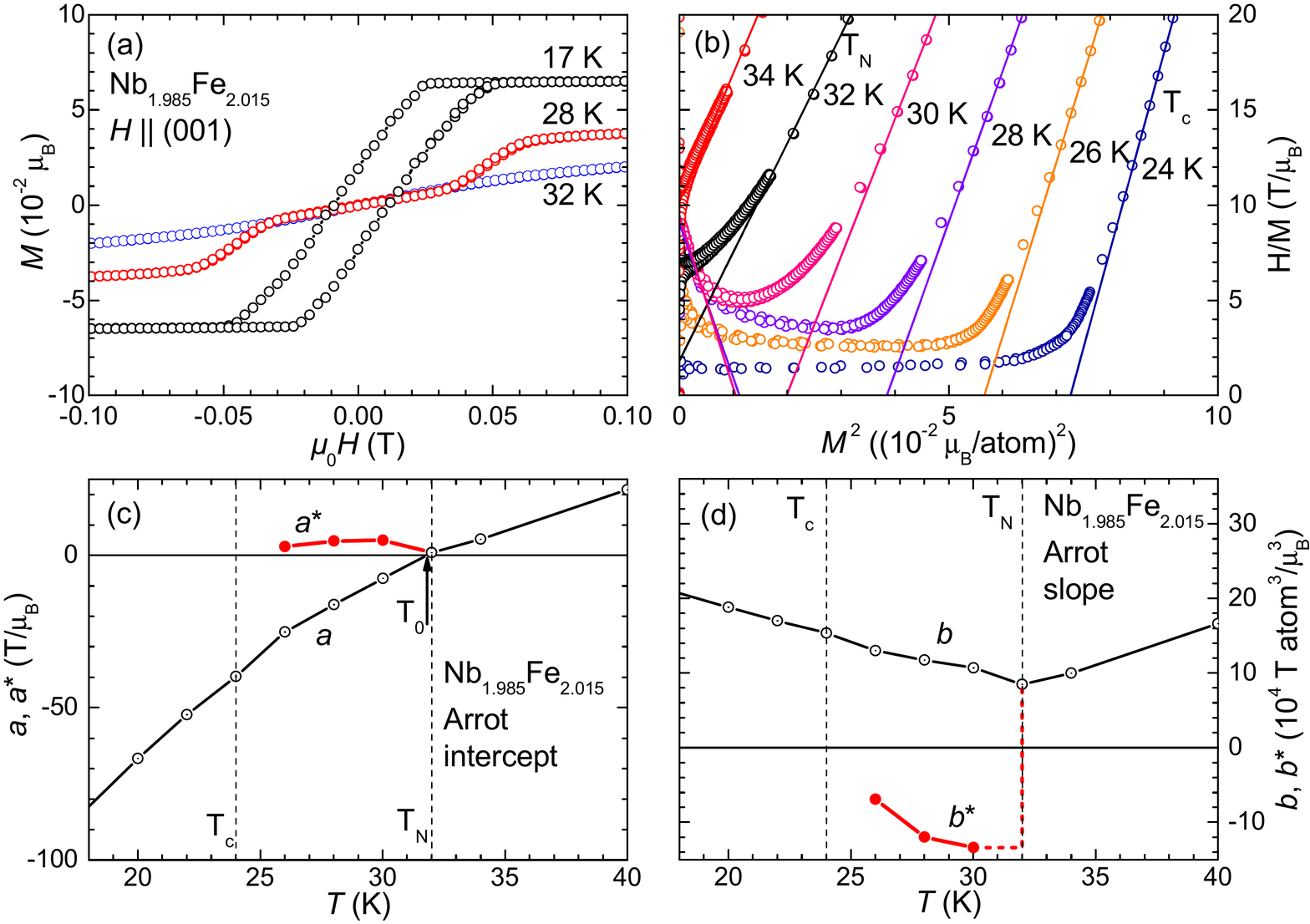}\\
	\includegraphics[width=.8\textwidth]{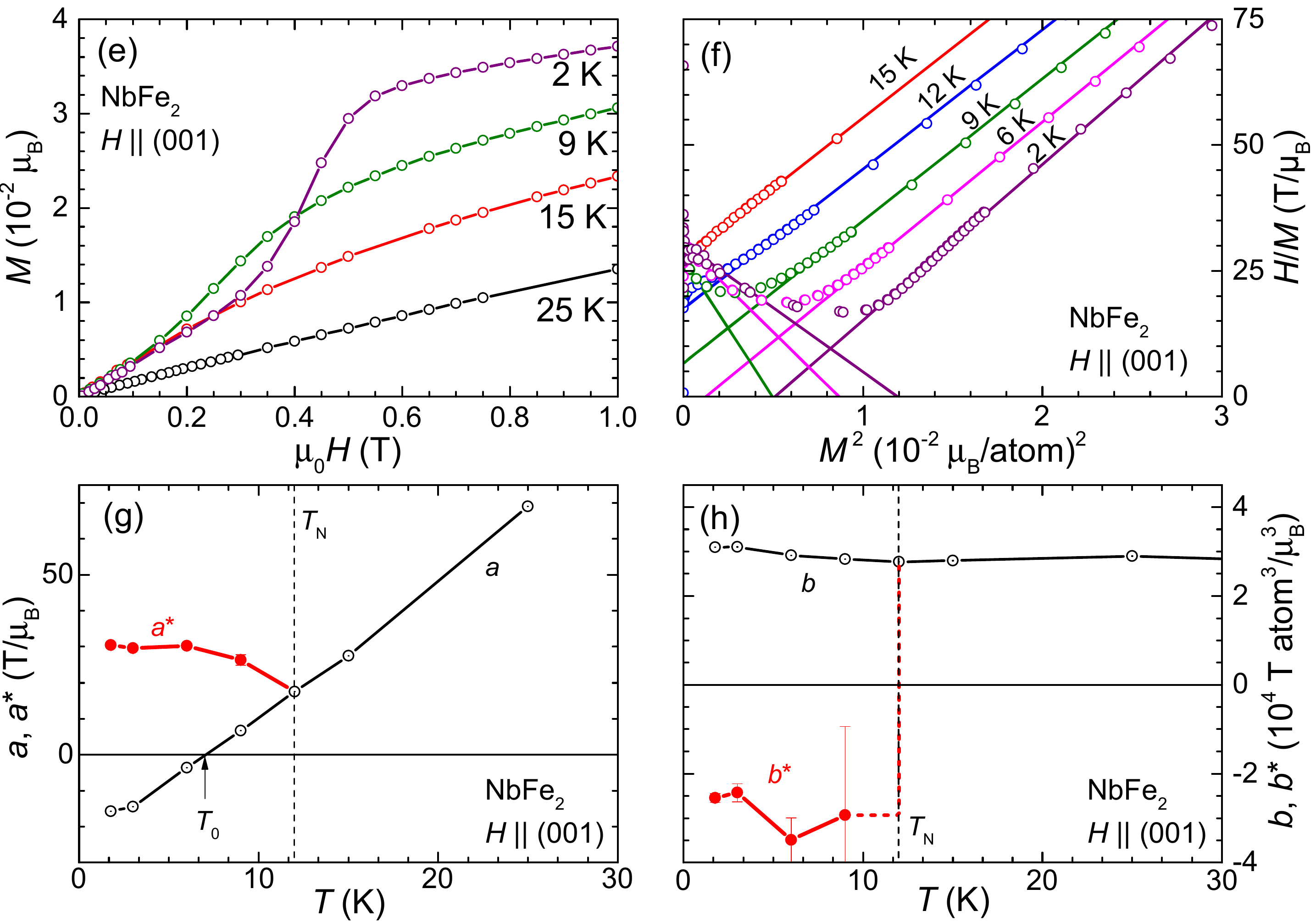}
\caption{\label{Arrott} DC magnetization isotherms in
  iron-rich \NbFeFe\ (a) and stoichiometric NbFe$_2$ (e) for fields along the     
	crystallographic $c$-direction. 	
	High and low-field straight-line fits to the Arrott plots of $H/M$ vs. $M^2$ (b, f) give intercepts $a$, \astar\ and slopes $b$, \bstar\ for
\NbFeFe\ (c, d) and \NbFe\ (g, h), enabling a direct comparison with a two order parameter Landau theory (see text).
%
}
\end{figure*}

Our high-quality single crystals 
show the previously established variation of the FM and 
SDW phases in a set of zero-field measurements (Fig.~\ref{ChiRes}).
In iron-rich \NbFeFe\ signatures of both FM and SDW transitions are seen 
at $\TFM \simeq \SI{24}{\kelvin}$ and $\TAFM \simeq \SI{32}{\kelvin}$, respectively 
as anomalies in 	the temperature dependent magnetic susceptibility $\chi(T)$, 
 linear thermal expansion $\dd L/(L_0 \dd T)$ and electrical resistivity $\rho(T)$ (cf.\ Fig.~\ref{ChiRes}). 
These signatures are consistent 
with 1st and 2nd order transitions at \TFM\ and \TAFM, respectively: 
The peak in $\chi(T)$ shows hysteresis at \TFM\ only, the thermal expansion shows 
a peak at \TFM\ and a kink at \TAFM, and the resistivity has 
a distinct kink at \TFM\ with hysteresis (Fig.~\ref{fig:Res_Hyst}), 
but only a much weaker anomaly is present 
in the derivative $\dd\rho/\dd T$ at \TAFM.
The FM state is unambiguously identified by remanent magnetization 
(Fig.~\ref{Arrott}(a)).

Stoichiometric \NbFe\ displays a single transition at $\TAFM=\SI{12}{\kelvin}$ with the characteristics of SDW order: A peak in $\chi(T)$ without hysteresis,
a kink in $\dd L/(L \dd T)$, and a weak enhancement in $\dd\rho/\dd T$ above the 
linear background from higher temperatures (Fig.~\ref{ChiRes}). Similarly, for niobium rich \NbFeNb, a single transition consistent with SDW order at $\TAFM=\SI{3}{\kelvin}$ can be inferred from the peak in $\chi(T)$ in Fig.~\ref{ChiRes}(a).

\begin{figure}[t]
\includegraphics[width=\columnwidth]{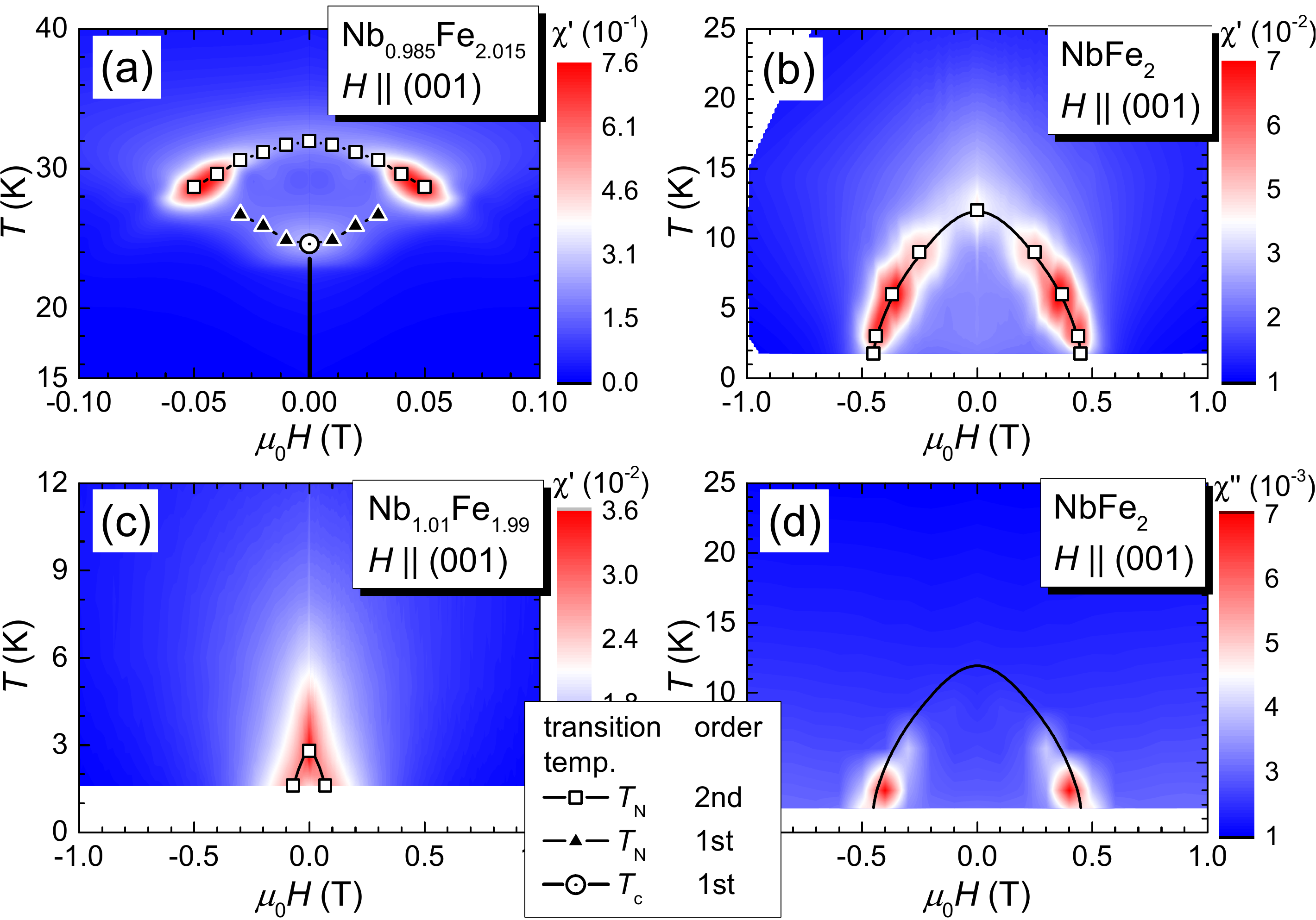}
  \caption{\label{Contours} Colour representation of the AC magnetic susceptibility $\chi(H,T)$ as a function of magnetic field and temperature for
    iron-rich Nb$_{0.985}$Fe$_{2.015}$ (a), stoichiometric NbFe$_2$ (b) and
    niobium-rich Nb$_{1.01}$Fe$_{1.99}$ (c). Points represent positions of the
    $\chi(T)$ maxima with (closed triangles) and without hysteresis (open squares) associated with \TAFM\ as well as of the maximum in zero field (open circle) associated with \TFM\ (see text). 
	(d) Imaginary part of the AC magnetic susceptibility $\chi^{\prime\prime}(H,T)$ for stoichiometric \NbFe. Solid line in (d) marks the phase boundary from (b).
		}
\end{figure}

The $H$-$T$ phase diagram is mapped \sven{for field parallel to the magnetic axis ($H\parallel c$)} 
using magnetic susceptibility $\chi(T, H)$ measurements for a series of \NbFey\ samples spanning the range from 
FM ground state via samples with SDW ground state 
to those in ultimate proximity to the SDW QCP (Fig.~\ref{Contours}).
In iron-rich \NbFeFe\ the maxima in $\chi(T)$ signalling 
the 1st order (\TFM) and 2nd order (\TAFM) transition shift 
to higher and lower temperature, respectively for increasing magnetic field.
The two signatures approach each other and eventually merge
at a critical field $\mu_0 \Hstar \simeq \SI{0.06}{\tesla}$ 
and a critical temperature $\Tstar \simeq \SI{28}{\kelvin}$. 
Only weak maxima reminiscent of crossovers are observed 
in $\chi(T)$ for fields above \Hstar.
Thus, the two lines of anomalies enclose the SDW phase which 
exists in the small parameter space between \TAFM\ and \TFM\ 
and for fields $H\leq\Hstar$ only.
At the critical point $(\Hstar,\Tstar)$, the susceptibility reaches 
highest values comparable to those observed in band ferromagnets 
with a 2nd order transition near the Curie temperature, 
such as ZrZn$_2$, if slight inhomogeneity and demagnetizing fields are present.  
We will later see that this enhanced susceptibility is 
expected within our model of competing order parameters
and marks the tricritical point at the transition 
from first order to second order at \Tstar\ \cite{Lawrie1984}.
The 1st and 2nd order nature of the low-temperature and high-temperature 
boundary of the SDW phase can be inferred from 
the presence and absence of hysteresis in the AC susceptibility and electrical resistivity
as detailed in Supplementary Material \ref{sec:SI:hyst}.

In stoichiometric \NbFe\ the 2nd order transition can similarly be followed through the $H$-$T$ phase diagram:
The maximum in $\chi(T)$ associated with \TAFM\ is shifted 
to lower temperatures upon increasing the magnetic field 
up to a critical field of $\HAFM \approx \SI{0.45}{\tesla}$. 
The line of anomalies separates out the low temperature, 
low field part of the $H$-$T$ phase diagram 
and suggests that in this region NbFe$_2$ forms a
distinct broken-symmetry state. 
This ``cap'' for the SDW phase is 
reminiscent of the upper part of the SDW phase in iron-rich \NbFeFe.
In fact, we show below that the same competing order parameter
model applies to both compositions and the phase diagram of stoichiometric
\NbFe\ resembles that of iron-rich \NbFeFe\ with the temperature axis shifted down by $\approx\SI{20}{\kelvin}$.

Whilst the SDW transition remains 2nd order in \NbFe\ for most of the phase boundary, we find signatures of a tricritical point at $(\Hstar=\SI{0.44}{\tesla},\Tstar=\SI{3}{\kelvin})$. Here, the susceptibility is strongly enhanced (Fig.~\ref{Contours}(b) \sven{and \ref{fig:OFZ28_chi3D}})  and a strong signal in the imaginary part $\chi^{\prime\prime}(\Hstar,\Tstar)$ is observed (Fig.~\ref{Contours}(d)) like in iron-rich \NbFeFe\ at the tricritical point (cf.\ Supplementary Information \ref{sec:SI:hyst}).  

\section{Discussion}
Identifying the lines of anomalies for $T<T^*$ as phase boundaries is
uncontroversial, because they are associated with hysteresis. The case
for a second order 'cap' linking the tricritical points at $(\pm\Hstar,\Tstar)$,
however, needs to be examined carefully and is reminiscent of the
situation in Sr$_3$Ru$_2$O$_7$ at high magnetic field. There, proof of
a broken symmetry state (as opposed to metamagnetic transition lines
ending in critical endpoints) came from thermal expansion and
thermodynamic data \cite{Rost2009}. In NbFe$_2$, further to earlier heat capacity
measurements on polycrystals \cite{moroni09}, strong 
support for the interpretation of the anomalies at \TAFM\ as phase
transition anomalies is provided by the thermal expansion shown in Fig.~\ref{ChiRes}(b), as well as the observation of finite ordered moments
within the SDW phase by ESR and $\mu$SR studies \cite{rauch15}.

Having established the presence of both the SDW and FM phase transitions
we seek a consistent description of the low temperature phase diagram of
\NbFey\ taking into account the proximity to both orders. 
At the most elementary level, this is done by postulating 
a Landau expansion of the free energy in terms
of two order parameters \cite{moriya77}:

\begin{equation}
\frac{F}{\mu_0} = \frac{a}{2} M^2 + \frac{b}{4} M^4 + \frac{\alpha}{2} P^2 +
\frac{\beta}{4} P^4 + \frac{\eta}{2} P^2 M^2 - H M
\label{eq:OP}
\end{equation}

Here, $M$ denotes the uniform magnetization, which couples linearly to
the applied magnetic field, whereas $P$ denotes a general second
order parameter, which does not couple directly to the applied field
but has a biquadratic coupling to the uniform magnetization. 
We associate the second order parameter $P$ with the SDW phase. The
phenomenological parameters $a$ and $b$ can be extracted directly from
magnetization measurements, for example $a = \chi^{-1}$ for $M=0$, but
the remaining parameters $\alpha$, $\beta$ and $\eta$ 
are more difficult to obtain. The theory can be formulated in terms of scalar
order parameters in isotropic materials, because the more complicated
coupling terms in a vector theory will constrain $M$ and $P$ either to
point in the same direction or at right angles to each other \cite{moriya77}. In anisotropic materials, the situation is in principle more
complicated, but as long as the field points along the easy axis
 as is the case for our studies of \NbFey\ here, the
scalar description remains adequate.

\begin{figure}%
\includegraphics[width=.5\columnwidth]{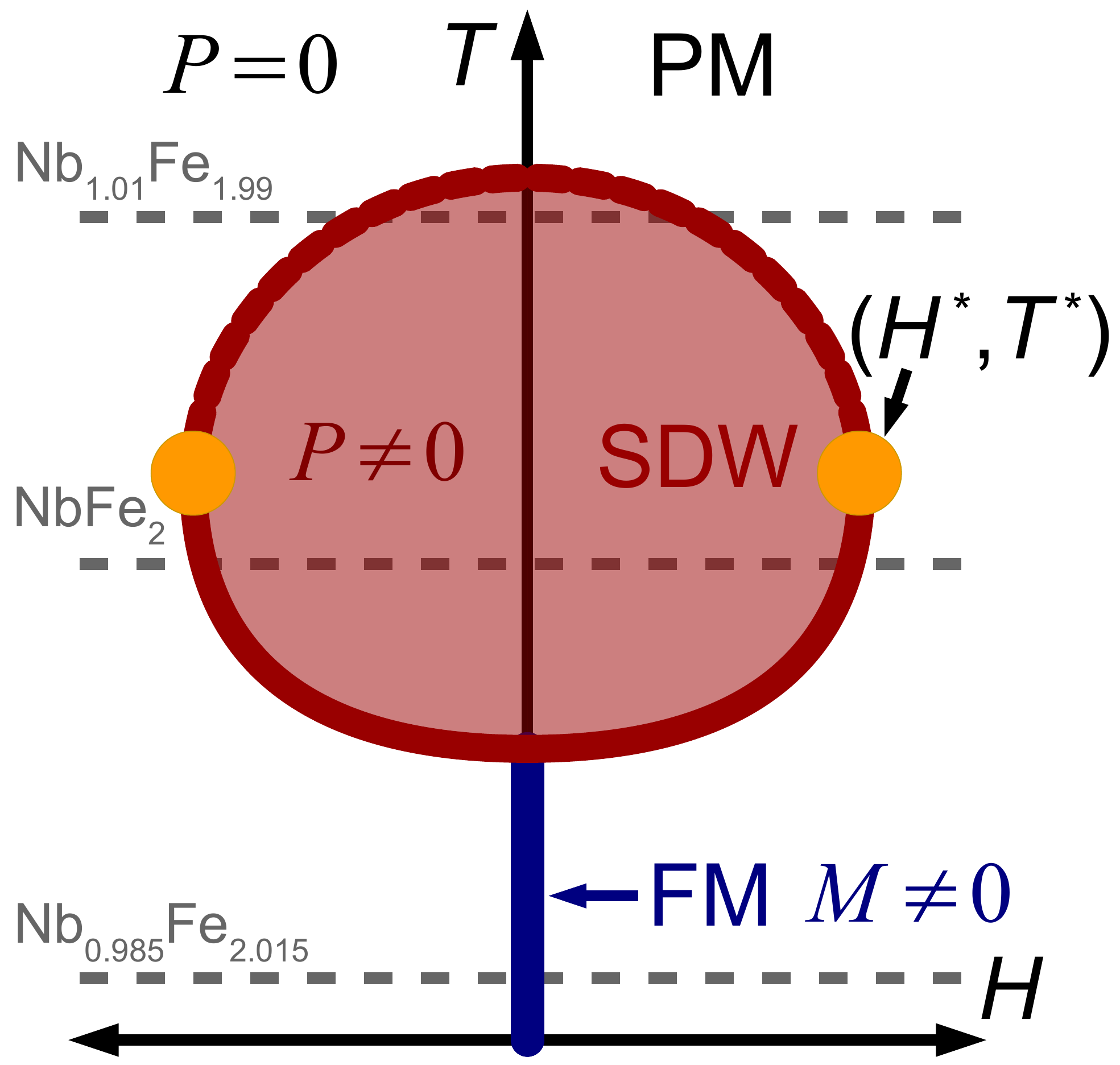}%
\caption{Schematic phase diagram based on the model free energy in Eqn.~(\ref{eq:OP}), as applied to \NbFey. Solid blue and red lines indicate the first order phase boundaries of the SDW and FM phase. Dashed red lines indicate the 2nd order phase boundary of the SDW phase at high temperatures. Orange circles mark the tricritical points. Horizontal gray dashed lines indicate zero-temperature assigned to the different samples of the composition series \NbFey.}%
\label{fig:PD}%
\end{figure}

In zero field the global free energy minima will
correspond to either a paramagnetic state $M=P=0$, or  one of the
possible magnetic states (i) $M=0$, $P \neq 0$; (ii) $M\neq 0$, $P=0$
or (iii) $M\neq 0$, $P\neq 0$, depending on the parameters $\{ a, b,
\alpha, \beta, \eta\}$. All prior observations in
polycrystalline NbFe$_2$, as well as our data on single crystals 
suggest that for $H=0$ the mixed phase $M \neq 0$, $P\neq 0$
does not occur in NbFe$_2$, and that on cooling the system will always
first develop the SDW order parameter ($P \neq 0$), before that is replaced 
by a uniform magnetization. This constrains $\alpha(T)$ to go
through zero at a higher temperature than $a(T)$.
The expected phase diagram for this case is illustrated in Fig.~\ref{fig:PD}.

We start by comparing the theoretical phase diagram with the 
observed behaviour in zero field: 
For $H=0$ and within the $P\neq 0$ state the free energy 
has its global minimum at $F_P =-\alpha^2/(4\beta)$ for $P^2 = -\alpha/\beta$. 
However, if the system
were to order uniformly, i.e. $M\neq 0$, the free energy would have a
minimum of $F_M = -a^2/(4 b)$ at $M^2 = -a/b$, and so a first order
transition from the $M=0$, $P\neq 0$ into the $M\neq0$, $P=0$ state
will occur at a temperature \TFM, below which $a^2/b > \alpha^2/\beta$. 
This sequence of a 2nd order transition into the SDW state
followed by a 1st order transition into the FM state exactly matches 
our observations in iron-rich \NbFeFe.

We now turn to the behaviour at finite field.
This is conveniently analysed at a  fundamental level 
by comparing theoretical and experimental Arrott 
plots of $M^2$ vs. $H/M$. Outside the SDW phase we have
\begin{align}
	\frac{H}{M} &= a  +b M^2 & \text{for} \quad P=0
\label{eq:ArrottFM}
\end{align}
allowing us to extract $a(T)$ as well as $b(T)$ 
as the intercept and slope, respectively.
Inside the SDW phase, i.e. for $P\neq0$,
this modifies the equation of state to 
$ H = (a+ \eta P^2) M + b M^3$
while at the same time minimizing the free energy with respect to $P$
gives $ P^2 = -\frac {\alpha}{\beta} - \frac {\eta}{\beta} M^2$ 
Substituting this into the equation of state results in a modified
expression for the Arrott plot within the SDW phase.
\begin{align}
	\frac{H}{M}&= \underbrace{\left(a-\frac{\alpha\eta}{\beta}\right)}_{\astar} + \underbrace{\left(b-\frac{\eta^2}{\beta}\right)}_{\bstar}M^2 &\text{for} \quad P\neq 0
\label{eq:ArrottAFM}
\end{align}
Thus, we expect a different slope and intercept in the SDW phase with $\astar(T)$ bifurcating from $a(T)$  and  $\bstar(T)$ jumping at \TAFM.

In Figs.~\ref{Arrott} (b) and (f) 
we analyse the high- and low-field parts of the Arrott plots for \NbFe\ and \NbFeFe\ 
according to eqs.~(\ref{eq:ArrottFM}) and (\ref{eq:ArrottAFM}). 
Indeed, the slope $\bstar(T)$ extracted from the low-field part of the Arrott plot changes discontinuously from a positive value $\bstar = b$ outside the SDW phase to a negative value $\bstar \neq b$ inside the SDW phase. 
The temperature dependence of the extracted parameters in Figs.~\ref{Arrott}(c),
(d), (g), and (h) agree with the expectations for 
a Landau theory for \NbFe, \NbFeFe, 
and \NbFeNb\ (not shown): 
(i) $b$ remains positive at all temperatures, 
(ii) $a$ has a strong temperature dependence,
(iii) $a$ and \astar\ bifurcate at \TAFM,
(iv) \bstar\ changes discontinuously at \TAFM.

The high-field Arrot-plot intercept $a(T)$ 
(Figs.~\ref{Arrott}(c) and (g)) crosses through 
zero at an intermediate temperature $\TC<\TAFM$ 
which indicates the underlying ferromagnetic instability. 
Ferromagnetism does not set in at \TC, because it has
been preempted by SDW order at \TAFM, but instead a first order
ferromagnetic transition occurs at a lower temperature $\TFM <  \TC$.

Considering next the shape of $M(H)$ isotherms 
on crossing the SDW phase boundary at constant 
$T$ yields two regimes within the two-order parameter model
(eqs.~\ref{eq:ArrottFM} and \ref{eq:ArrottAFM}) \cite{moriya77}. 
At low temperatures $M(H)$ is predicted to evolve 
discontinuously through the phase boundary whilst at high
temperatures $M(H)$ evolves continuously.
This implies that the phase boundary between the SDW state and
the finite field paramagnetic state is expected 
1st order and 2nd order at low and high temperatures
with the two regimes separated by a  tricritical point of divergent 
susceptibility at the  maximum critical field of the SDW phase 
(cf.\ Fig.~\ref{fig:PD}). This separation into a 1st and
2nd order regime of the SDW phase boundary entirely matches
our observations in iron-rich \NbFeFe\ including the presence
of a tricritical point with strongly enhanced 
susceptibility as discussed above (Fig.~\ref{Contours}).
For stoichiometric \NbFe\ and niobium-rich \NbFeNb\ 
only parts of the predicted phase diagram are accessible
as indicated by the horizontal lines in Fig.~\ref{fig:PD}:
In \NbFe\ we observe the 2nd order upper phase boundary 
of the SDW phase and the tricritical point 
with a cut-off just below \Tstar\
whilst in \NbFeNb\ only the top part of the SDW phase is 
observed with a cut-off just below \TAFM.

\begin{figure}[t]
\includegraphics[width=\columnwidth]{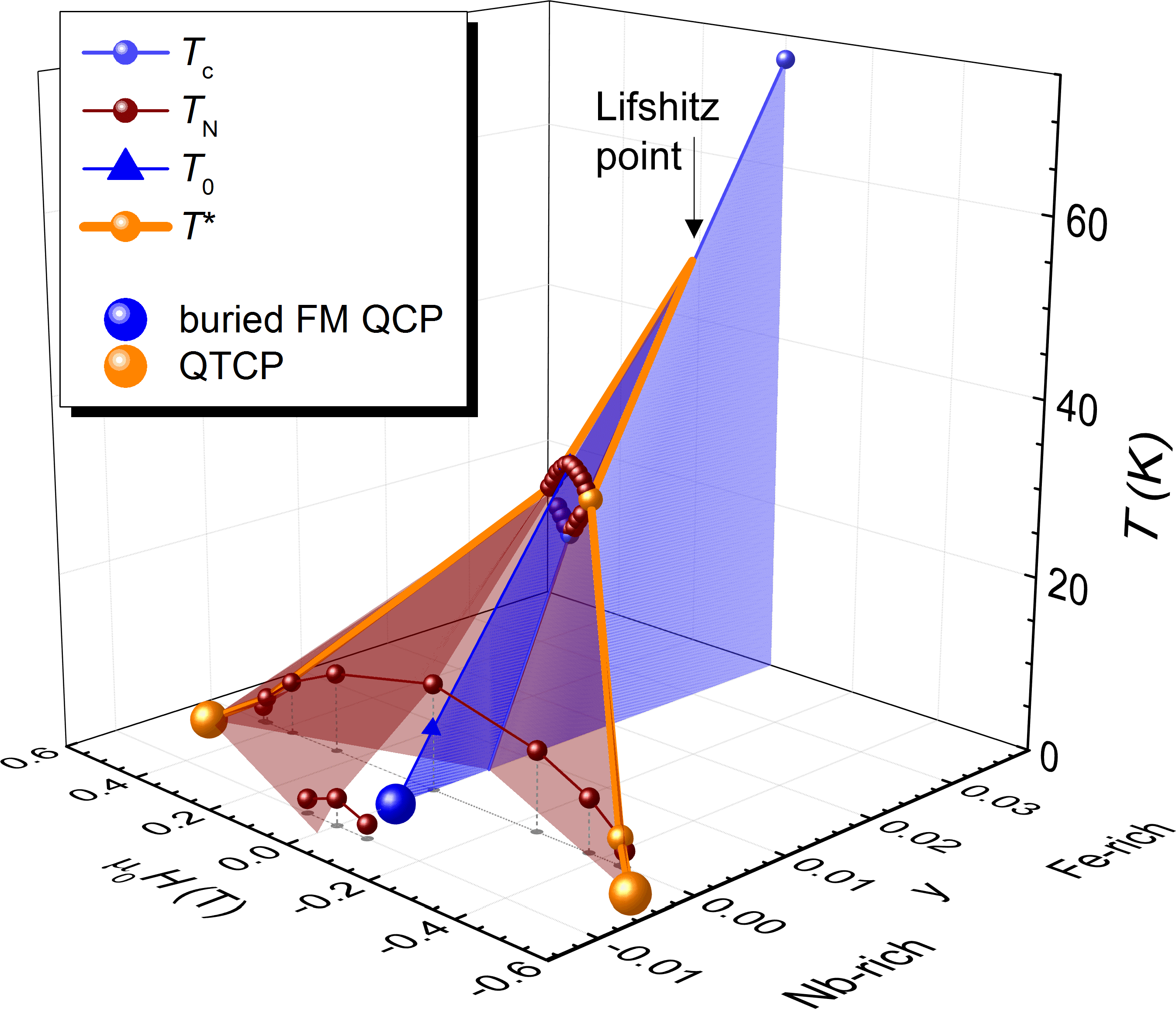}
  				
  \caption{\label{PhaseDia} Overall composition--magnetic
    field--temperature phase diagram for the \NbFey\ system. The 
    underlying ferromagnetic transition temperature \TC\ is extracted from
    $a(\TC) = 0$ (cf.\ Figs.~\ref{Arrott}(c) and (g)). 
		The phase boundaries of the SDW and FM
		phase are obtained from the magnetization and susceptibility measurements
		as shown in Figs.~\ref{ChiRes}, \ref{Contours}, and \ref{Arrott}.
		The position of the avoided ferromagnetic QCP (blue ball) and the 
		QTCPs (orange ball) are highlighted.
		}
\end{figure}

Our findings are summarized in the global $H$-$T$-$y$ phase diagram 
in Fig.~\ref{PhaseDia}, which illustrates that the accessible part
of the schematic phase diagram (Fig.~\ref{fig:PD}) shrinks with 
increasing Nb content.
The global phase diagram also illustrates the decrease of \TC\ 
as the composition is varied from Fe-rich towards Nb-rich 
and that \TC\ extrapolates to zero temperature at $y\approx- 0.004$ and $H=0$.
This marks the avoided FM QCP buried inside a dome formed by the SDW phase.
The intrinsic tendency of clean metallic systems 
to avoid a FM QCP either by changing the nature 
of the phase transition from 2nd order to 1st order 
or by developing competing SDW magnetic order 
has long been noted \cite{belitz97,Vojta99, Brando2016, Conduit2009}. 
Our observation of emergent SDW order 
enveloping the preempted FM QCP represents 
the first example of the latter scenario 
among itinerant magnets, complementing 
the recent report of emergent helical order 
in the local moment system PrPtAl \cite{Abdul-Jabbar2015}.

In addition to the buried FM QCP 
the global phase diagram (Fig.~\ref{PhaseDia}) 
reveals another important insight, 
the presence of quantum tricritical points (QTCPs):
Finite-temperature tricritical points have been located
for iron-rich \NbFeFe\ at $\Tstar \approx \SI{28}{\kelvin}$ 
and for stoichiometric \NbFe\ at $\Tstar\approx\SI{3}{\kelvin}$.
This demonstrates that the tricritical points can be 
suppressed to zero temperature leading to QTCPs. 
Based on a smooth interpolation we estimate 
the location of the QTCPs 
at ($y\simeq -\SI{0.003}{}, \pm\mu_0 H\simeq\SI{0.5}{\tesla},T=0$).

A divergent uniform susceptibility is not only expected 
within the two-order-parameter description above \cite{moriya77}
but also within a self-consistent spin-fluctuation theory for 
antiferromagnetic order in itinerant systems \cite{Misawa2008}.
The divergent uniform susceptibility near the QTCP causes strong 
FM  fluctuations which may contribute to the 
logarithmic divergence of the specific heat observed near the SDW QCP at 
($y\simeq -\SI{0.01}{},  H=0, T=0$) \cite{brando08}. 
Indeed, recent theoretical work suggests that the finite 
temperature behaviour at an SDW QCP may be dominated by  
FM fluctuations of a nearby FM QCP above a 
crossover temperature that is different 
for different physical quantities \cite{Oliver15}. 
In \NbFe, we have a QTCP with FM fluctuations. 
At the nearby SDW QCP these FM fluctuations
can produce $C/T\propto\log(T)$ above 
a low-lying crossover temperature specific
to the heat capacity, whereas the corresponding crossover 
for resistivity may be higher, such that the signatures 
of SDW QCP are retained in $\rho(T)$ at low $T$. 

Analysing our experimental results in newly available 
single crystals of the band magnet \NbFe\ 
and its iron-rich composition series in
terms of a simple but powerful two order-parameter 
Landau theory has brought to light 
a new generic phase diagram for the vicinity of the
FM QCP in clean metallic systems: 
(i) the FM QCP is enveloped by a dome 
of emergent SDW order, 
(ii) divergent $\chi$ is shifted to tricritical
points at finite field, (iii) the line of 
tricritical points terminates at finite field 
at zero temperature, generating a QTCP. The coincidence of
multiple phase boundaries and critical points 
may underlie the experimental observation 
that $C/T$ follows the $\log(T)$ behaviour characteristic
of a FM QCP, whereas $\rho(T)$ displays the $T^{3/2}$ 
power law expected near an SDW QCP \cite{brando08}.

The identification of generic QTCPs in \NbFe\ 
opens up the new phenomenon of quantum tricriticality 
for experimental studies in a whole class of 
systems with buried or avoided FM QCP.
\sven{This provides a fresh perspective on other materials with the
same universality, including prototypical heavy-fermion materials 
\cite{Misawa2008,Misawa2009},
in which multiple and competing low-energy scales have 
in the past prevented the detection of a QTCP and 
obscured the investigation of its consequences.}

%

\section{Materials and Methods}
Samples from the composition series 
\NbFey\ with $-\SI{0.005}{} < y < \SI{0.015}{}$ were grown in an
adapted, UHV-compatible mirror furnace from polycrystalline ingots
prepared by radio-frequency induction melting, as described
previously \cite{friedemann13}. Single crystal grains were selected and oriented
by x-ray and neutron diffraction. Magnetic and resistivity measurements 
were carried out on a Quantum Design PPMS.
Thermal expansion measurements made use of a
custom-designed dilatometry insert for the PPMS \cite{Kuchler12}.

\section{Data Availability}
All data needed to evaluate the conclusions in the paper are present in the
paper, the Supplementary Materials and the Data repository at the University of Cambridge and can be download from https://www.repository.cam.ac.uk/handle/xxxx/xxxxx. Additional data related to this paper may be requested from the authors.


\section{Acknowledgement} We thank G. G. Lonzarich and
P. Niklowitz for helpful discussions. This work was supported by
the EPSRC UK under grant No EP/K012894, the Alexander-van-Humboldt foundation, FOR 960 Quantum Phase Transitions, and Transregio 80 (TRR80).


%

\newpage
\begin{center}
	\textbf{ \Large Supplementary Materials to \\ ``Buried ferromagnetic quantum critical point in single-crystal \NbFe ''}
\end{center}
\newpage
%
%
\renewcommand{\textfraction}{0.07}

\renewcommand{\thefigure}{S\arabic{figure}}
\renewcommand{\thetable}{S\arabic{table}}
\renewcommand{\thesection}{\Roman{section}}
\setcounter{figure}{0}
\setcounter{table}{0}
\setcounter{section}{0}
%
%
\section{Phase transition characteristics in \protect \NbFeFe}
\label{sec:SI:hyst}

\begin{figure}%
\includegraphics[width=.75\columnwidth]{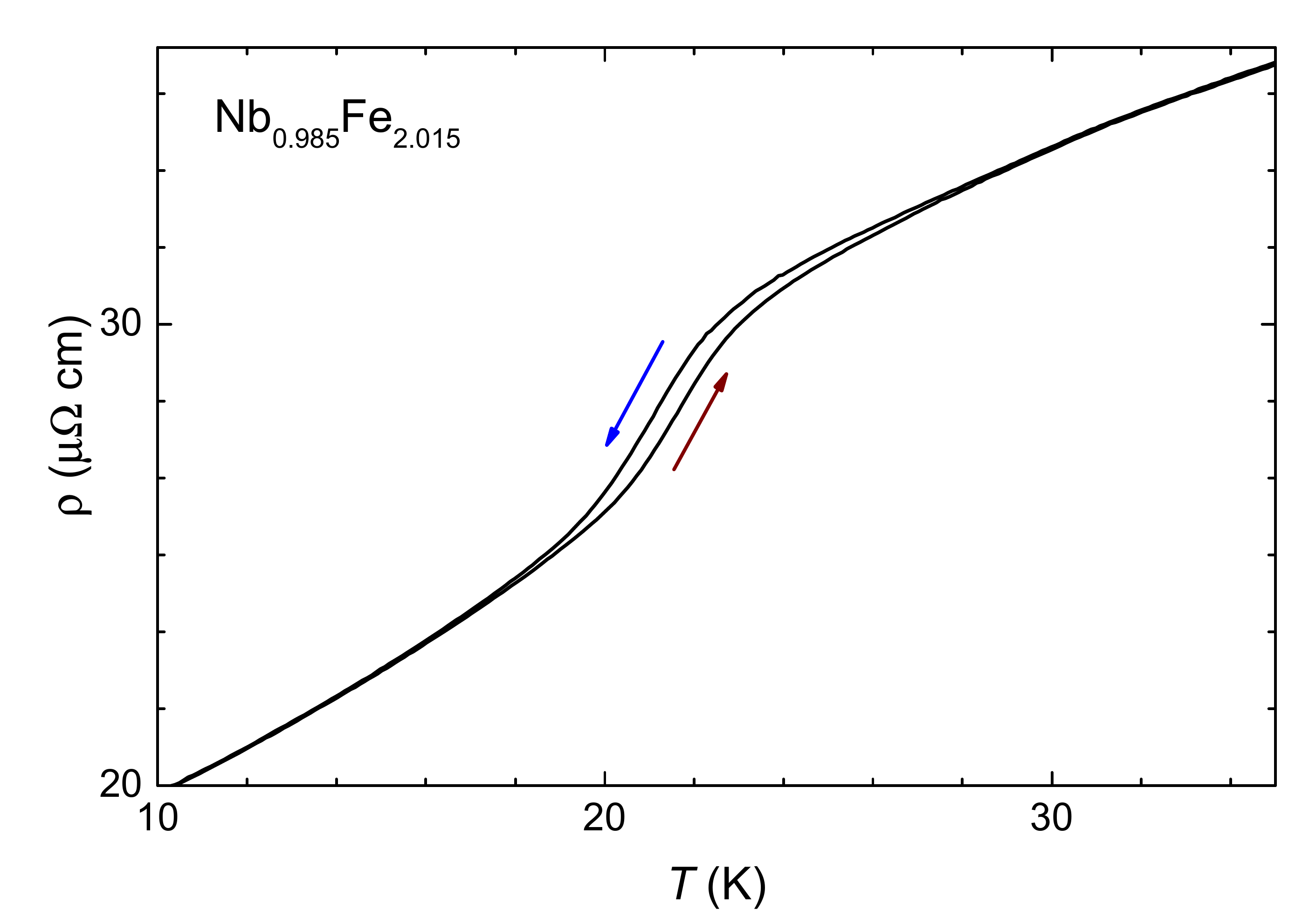}%
\caption{Hysteresis at the FM transition in the resistivity for \NbFeFe. Arrows indicate the direction of the temperature sweep.}%
\label{fig:Res_Hyst}%
\end{figure}

The zero-field transition into the ferromagnetic state of \NbFeFe\ shows
clear hysteresis as evident from resistivity measurements in Fig.~\ref{fig:Res_Hyst} and the susceptibility (Fig.~\ref{ChiRes}(a)). This implies a first-order transition
at \TFM.

\begin{figure}%
\includegraphics[width=\columnwidth]{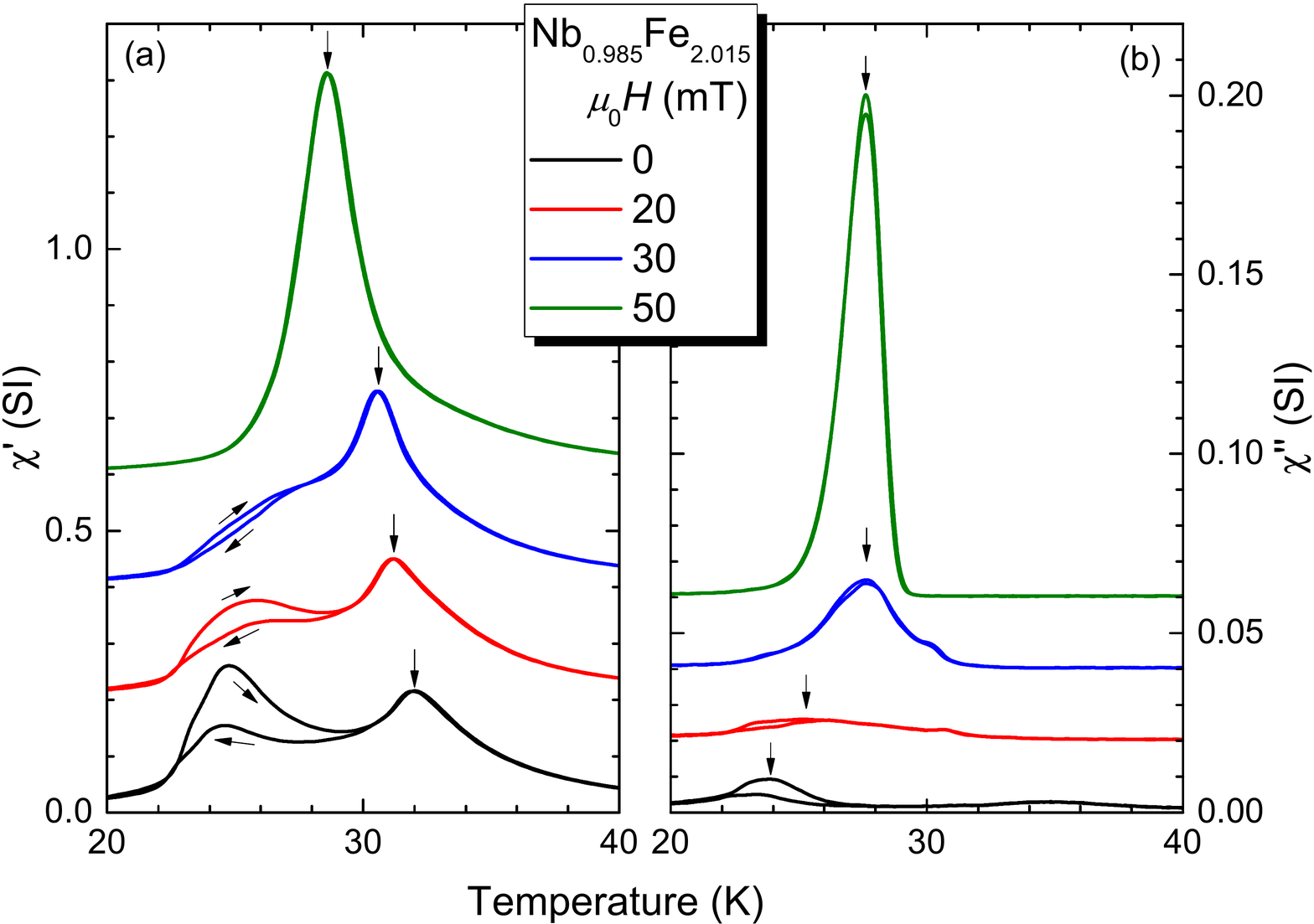}%
\caption{Hysteresis in temperature sweeps of the AC susceptibility real $\chi^{\prime}(T)$ and imaginary part $\chi^{\prime\prime}(T)$. Vertical arrows in (a) and (b) indicate the upper and lower transition of the SDW state, respectively. Arrows around the lower transition in (a) indicate the direction of sweeping the temperature. Data are offset by $0.2$ (a) and $0.02$ (b) for clarity.}%
\label{fig:chi_Hyst}%
\end{figure}

In finite field, hysteresis is present below \Tstar\ at the low-temperature transition of the SDW phase only as observed in the susceptibility (Fig.~\ref{fig:chi_Hyst}(a)). 
This implies a first order transition for $T<\Tstar$ and a second order transition
for $T>\Tstar$ with a tricritical point at $(\Hstar,\Tstar)$.
In addition, we find a peak in the imaginary part of the AC
susceptibility $\chi^{\prime\prime}(T)$ along the low-temperature boundary 
of the SDW phase (Fig.~\ref{fig:chi_Hyst}(b)). Indeed this peak becomes more 
pronounced on the approach of \Hstar\ indicative of the 1st order
becoming weaker and thus promoting strong dissipation. 
This trend culminates in a strong enhancement of $\chi^{\prime\prime}(T)$
right at \Hstar\ for temperatures below \Tstar\ and suggests ultimate 
proximity to a tricritical point with fluctuations in the 
uniform susceptibility.

\begin{figure}%
\includegraphics[width=\columnwidth]{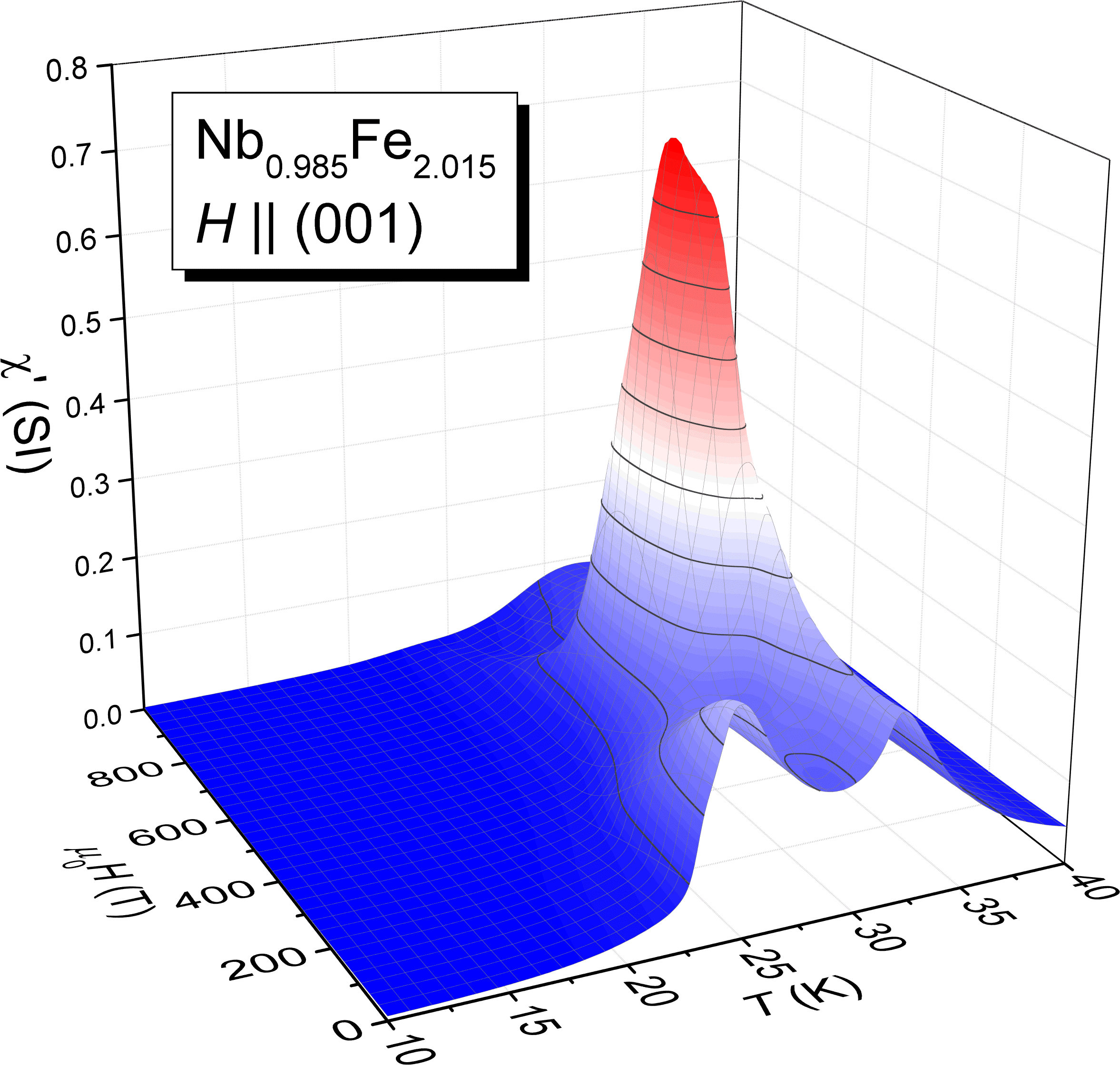}%
\caption{\sven{Magnetic susceptibility map for iron-rich \NbFeFe. The three-dimensional representation was constructed from warming temperature sweeps after field cooling.}}%
\label{fig:OFZ28_chi3D}%
\end{figure}

\sven{Figure \ref{fig:OFZ28_chi3D} highlights the enhancement of the uniform susceptibility at $(\Hstar,\Tstar)$. These divergent fluctuations at zero wavevector together with divergent fluctuations at finite wavevector which are implied by the 2nd order nature of the SDW transition at \Tstar\ characterise the tricritical point at $(\Hstar,\Tstar)$.}

\end{document}